\documentclass[manuscript]{acmart}

\AtBeginDocument{%
  \providecommand\BibTeX{{%
    \normalfont B\kern-0.5em{\scshape i\kern-0.25em b}\kern-0.8em\TeX}}}

\usepackage{multirow}
\usepackage{graphicx}
\usepackage{xcolor}
\definecolor{secondary}{gray}{0.75}
\copyrightyear{2023}
\acmYear{2023}
\setcopyright{rightsretained}
\acmConference[TAS '23]{First International Symposium on Trustworthy
Autonomous Systems}{July 11--12, 2023}{Edinburgh, United Kingdom}
\acmBooktitle{First International Symposium on Trustworthy Autonomous
Systems (TAS '23), July 11--12, 2023, Edinburgh, United
Kingdom}\acmDOI{10.1145/3597512.3599697}
\acmISBN{979-8-4007-0734-6/23/07}

\begin{document}

\title{RE-centric Recommendations for the Development of Trustworthy(er) Autonomous Systems}

\author{Krishna Ronanki}
\affiliation{%
  \institution{University of Gothenburg}
  \streetaddress{Hörselgången 5}
  \city{Göteborg}
  \country{Sweden}
  \postcode{41756}
}
\email{krishna.ronanki@gu.se}

\author{Beatriz Cabrero-Daniel}
\affiliation{%
  \institution{University of Gothenburg}
  \streetaddress{Hörselgången 5}
  \city{Gothenburg}
  \country{Sweden}
  \postcode{41756}
}
\email{beatriz.cabrero-daniel@gu.se}

\author{Jennifer Horkoff}
\affiliation{%
  \institution{Chalmers | University of Gothenburg}
  \streetaddress{Hörselgången 5}
  \city{Gothenburg}
  \country{Sweden}
  \postcode{41756}
}
\email{jennifer.horkoff@gu.se}

\author{Christian Berger}
\affiliation{%
  \institution{University of Gothenburg}
  \streetaddress{Hörselgången 5}
  \city{Gothenburg}
  \country{Sweden}
  \postcode{41756}
}
\email{christian.berger@gu.se}

\begin{abstract}
  Complying with the EU AI Act (AIA) guidelines while developing and implementing AI systems will soon be mandatory within the EU. However, practitioners lack actionable instructions to operationalise ethics during AI systems development. A literature review of different ethical guidelines revealed inconsistencies in the principles addressed and the terminology used to describe them. Furthermore, requirements engineering (RE), which is identified to foster trustworthiness in the AI development process from the early stages was observed to be absent in a lot of frameworks that support the development of ethical and trustworthy AI. This incongruous phrasing combined with a lack of concrete development practices makes trustworthy AI development harder. To address these concerns, we formulated a comparison table for the terminology used and the coverage of the ethical AI principles in major ethical AI guidelines. We then examined the applicability of ethical AI development frameworks for performing effective RE during the development of trustworthy AI systems. A tertiary review and meta-analysis of literature discussing ethical AI frameworks revealed their limitations when developing trustworthy AI. Based on our findings, we propose recommendations to address such limitations during the development of trustworthy AI. 
\end{abstract}


\maketitle

\section{Introduction}

Due to enhanced access to large volumes of data generated all the over the world, we observe an accelerated use of artificial intelligence (AI) in our daily lives~\cite{iyer2021ai,kaur2022trustworthy}. A prominent aspect of AI is its ability to engage in its tasks autonomously~\cite{chesterman2020artificial}. As the adoption of AI across various fields, such as healthcare, transportation, and education, is growing, the need for AI to be trustworthy is also increasing every day. Moreover, Autonomous Systems (AS) are expected to exhibit ``broad intelligence'' while collaborating with human agents in a harmonious way by processing data in order to adapt to complex and unpredictable environments~\cite{sifakis2022trustworthy}. It is therefore more crucial than ever for these AS to be not only technically robust but also compliant with relevant development guidelines and regulations~\cite{kaur2022trustworthy}.

Trustworthy AI can be defined as a conceptual framework that ensures that the development and implementation of technically and socially robust AI systems adhere to all the applicable laws and regulations, and conform to general ethical principles~\cite{aihleg}. The concept of trustworthy AI can foster the realization of AI's full potential by establishing trust in the development and implementation of AI to maximize its benefits and reduce its risks at the same time~\cite{thiebes2021trustworthy}. In AS, multiple interrelated AI systems are typically incorporated to effectively automate complex tasks while seamlessly interacting with their users. A failure in any of these tasks could severely harm humans; therefore, trustworthy AI systems should play an important role in the development of AS~\cite{fernandez2021trustworthy}.

Often referred to as ``soft-law'', ethical AI guidelines provide recommendations for AI developers without necessarily being enforceable legislation~\cite{sossin2002hard}. Ethical AI guidelines aim to ensure that AI systems respect human rights and local ethical values to promote users' trust in AI, and to guide decision-making tasks related to the design and development processes of ethically compliant AI systems~\cite{campbell2000legal}. While these ethical guidelines are important, developers often lack any clear actionable instructions on how to practically implement them~\cite{mittelstadt2019principles}. Despite the establishment of numerous ethical AI principles within these AI guidelines, the vague terminology has not been helpful to data scientists, machine learning (ML) engineers, and designers who are essential in operationalising these principles to identify and address ethical concerns in practice~\cite{10.1145/3430368}. 

Addressing these issues early on can reduce the risk of unnecessary additional effort later in the development process such as verifying and validating the system's compliance with the ethical standards~\cite{vakkuri2021eccola}. Requirements Engineering (RE) involves ``\textit{all life cycle activities devoted to the identification of user requirements, analysis of the requirements to derive additional requirements, documentation of the requirements as a specification, and validation of the documented requirements against user needs, as well as processes that support these activities}''~\cite{berenbach2009software}. RE is considered a critical juncture of the interplay between ethics and technology~\cite{9218169}. The RE process at the beginning of a product development life cycle fosters increased communication and collaboration between various stakeholders, offering opportunities to discuss ethical concerns~\cite{kostova2020interplay} and incorporate them into the development process in a concrete manner. Ethical AI creation frameworks aim to guide the design, development, evaluation, operation and retirement~\cite{floridi2022capai} of AI systems. 

Many efforts are being devoted to adapting the RE process to the aforementioned AI guidelines. Several frameworks~\cite{dignum2019humane, baker2021taii,vakkuri2021eccola} have attempted to tackle the issue of operationalising and formalising ethical AI development guidelines, to turn vaguely coined guidelines into precise and measurable practices. But the focus on the RE processes is lacking for incorporating high-level ethical concerns (e.g., trustworthiness) in the early stages of the development of trustworthy AI, making it an open challenge~\cite{amaral2021trustworthiness}. Given the value RE can provide in the design phase of AI development life cyle~\cite{kostova2020interplay} and the expected emergence of new domestic and international laws applicable to AI development within the European Union (EU), this study aims to answer the following research questions:

 \begin{itemize}
    \item \textbf{RQ1:} What are the differences between the EU AI Act and other ethical AI guidelines?
    \item \textbf{RQ2:} What are the limitations of existing ethical AI development frameworks while developing trustworthy AI systems? 
    \item \textbf{RQ3:} What recommendations can be proposed to be implemented in the early stages of AI development to address the identified limitations and promote trustworthy AI?
\end{itemize}

\section{Related Work}

 This section includes a review of existing ethical AI guidelines, frameworks, and a reference development life cycle model for trustworthy AI. Additionally, we motivate the role of RE activities in achieving trustworthiness in AI systems by providing insights into the current state of the art in the field by identifying gaps in existing research to be addressed by our study. 

\subsection{Ethical AI guidelines}\label{sec:aiguidelines}

The IEEE Ethically Aligned Design (EAD) framework aims to promote considerations of ethical values and principles in the development and deployment of emerging technologies including but not limited to AS and AI. Their recommendations cover eight major values: \textit{Human Rights}, \textit{Justice}, and \textit{Responsibility},\textit{ Design Principles} for aligning AI systems with these values, guidelines for \textit{Responsible Data Management}, \textit{User-centred Design} and \textit{Governance} of AI systems~\cite{ieee}. 

International organisations such as UNESCO and IEEE have proposed AI guidelines with principles for responsible and ethical AI development and implementation. Similarly, governments from the United States of America, the European Union, the United Kingdom, Japan, Canada, Iceland, Norway, the United Arab Emirates, India, Singapore, South Korea and Australia, which jointly represent 31.9\% of the total world population~\cite{jobin2019global}, drafted or provided AI guidelines. China has also recently released a regulation for Generative AI~\cite{chinaAI}. Such guidelines also provide recommendations for international cooperation strategies and call for a dialogue between policy-makers and other stakeholders in the development and use of AI. 

The Japanese government established an expert group that assists in creating a national strategy on how AI principles should be implemented. The expert group focuses on principles and values like \textit{Human-centricity}, \textit{Education/Literacy}, \textit{Privacy}, \textit{Security}, \textit{Competitive fairness}, \textit{Accountability}, \textit{Transparency}, \textit{Fairness} and \textit{Innovation}~\cite{JapanAI}. The Japanese government also implemented initiatives and funding programs to support the development and adoption of AI technologies in industry and research.

UNESCO's recommendations on the Ethics of Artificial Intelligence~\cite{unesco} emphasise the need to understand and respect human rights and cultural diversity when developing AI systems, promoting \textit{Inclusiveness and Social Responsibility}, and ensuring \textit{Transparency} and \textit{Accountability} in AI decision-making. UNESCO aims to foster social progress and sustainable development and to protect human rights by promoting an ethical and human-centred approach to the development and use of AI. 

The Organisation for Economic Cooperation and Development (OECD), an intergovernmental organisation with 38 member countries representing 17.7\% of the global population, published a set of principles and recommendations for the responsible development and use of Artificial Intelligence (AI)~\cite{OECD}. The OECD AI Guidelines provide a framework for the responsible development and use of AI, with a focus on promoting trust and innovation in AI. The guidelines set out five principles for responsible stewardship of trustworthy AI, including \textit{Inclusive Growth and Sustainable Development}, \textit{Human-centred Values and Fairness}, \textit{Transparency and Explainability}, \textit{Robustness}, \textit{Security and Safety}, and \textit{Accountability}. In addition to these principles, the guidelines also provide recommendations for policy-makers on national policies and international cooperation for trustworthy AI, as well as provisions for the development of metrics to measure AI research, development, and deployment and for building a foundation of evidence to assess progress in implementation.

The Ethics Guidelines developed by European Commission's (EC) High-level expert group on artificial intelligence (AI HLEG) have identified four ethical principles that form the basis for TAI which include \textit{Respect for Human Autonomy}, \textit{Prevention of Harm}, \textit{Fairness}, and \textit{Explicability}~\cite{aihleg,AIact}. But in order to achieve TAI in practice, they believe that there are seven key requirements that must be continuously assessed and managed throughout the entire lifecycle of an AI system:\textit{ Human Agency and Oversight}, \textit{Technical Robustness and Safety}, \textit{Privacy and Data Governance}, \textit{Transparency}, \textit{Diversity, Non-discrimination, and Fairness}, \textit{Societal and Environmental Well-being}, and \textit{Accountability}.

Existing ethical AI frameworks may also have unnoticeable limitations due to the lack of cultural diversity and representation in their development. Inappropriate representation or exclusion of cultural heritage can lead to serious consequences such as discriminatory outcomes, perpetuation of stereotypes, and even harm to certain groups. Therefore, it is important for the development of AI systems to consider cultural diversity and representation in the RE phase to ensure that the ethical AI frameworks themselves are inclusive, fair, and trustworthy as well. The responsible research and innovation (RRI) framework~\cite{RRI} is an effective way to ensure that these factors are considered. The RRI framework encourages researchers and practitioners to think critically about potentially controversial topics, identify ethical and moral components, anticipate any controversial implications, and address issues of trust and social acceptability. Additionally, the RRI framework emphasizes the importance of broader interactions and responsible innovation engagement to influence research direction and trajectory.

\subsection{Ethical AI Frameworks} \label{sec:frameworks}

Crowley et al.~\cite{crowley2019toward} present the principles of responsible AI, which include \textit{Adaptability}, \textit{Responsibility}, and \textit{Transparency}. They propose a design methodology called Design for Values (DFV) to guide the development of responsible AI systems. This methodology includes steps such as aligning system goals with human values, using explicit interpretation mechanisms, specifying reasoning methods for ethical deliberation, specifying governance mechanisms for responsibility, ensuring openness, and promoting informed participation of all stakeholders. They also discuss how these principles can be integrated into a system development life cycle framework and the legal issues surrounding legal protection by design. The DFV methodology offers a structured approach for formulating requirements that take into account multiple normative interpretations of a value. However, the authors note that personal views may influence the design decision-making process when using DFV. To mitigate bias and ensure a more systematic approach, it is important to consider ways to reduce the room for misinterpretation when translating a value to a norm, which may impact the formulated functionality and requirements associated with it.

Hallensleben et al.~\cite{vcio} present a framework for incorporating ethical principles into the design, implementation, and evaluation of AI systems. Their groups are composed of experts in fields such as computer science, philosophy, and social sciences, and they are part of the AI Ethics Impact Group (AIEI Group). The framework provides concrete guidance for decision-makers in organizations developing and using AI on how to incorporate values into algorithmic decision-making and how to measure the fulfilment of values. It also offers practical ways of monitoring ethically relevant system characteristics as a basis for policymakers, regulators, oversight bodies, watchdog organisations, and standards development organizations. The ultimate goal is to achieve better control, oversight, and comparability of AI systems and to inform choices by citizens and consumers. 

Baker-Brunnbauer proposed the TAII Framework~\cite{baker2021taii}, a step-by-step approach for organizations to implement trustworthy AI by identifying and addressing ethical issues and dependencies throughout the AI system's value chain. The framework includes twelve stages that will be continually revisited throughout the entire life cycle of the AI system. It begins with creating an AI system brief overview, outlining the purpose, use case, and input data of the AI system. The framework also recommends establishing an internal ethics board, using visualized business models, and considering existing regulations and standards for the specific AI system. The success of the TAII process depends on the priority given to AI ethics, the allocated resources, and the commitment of stakeholders. 

A major observation of the above-mentioned ethical AI development frameworks is the lack of emphasis on systematic RE processes like requirements elicitation, analysis, specification, and management of requirements. These observations are further strengthened by the work of Floridi et al.~\cite{floridi2022capai}, who define five key stages in the life cycle of an AI system. 1. \textit{Design}, 2. \textit{Development}, 3. \textit{Evaluation}, 4. \textit{Operation}, and 5. \textit{Retirement} across three different environments, the \textit{training environment}, the \textit{testing environment}, and the \textit{production environment}. The design stage of this proposed life cycle model is dedicated to performing tasks including the \textit{formulation of use cases}, \textit{assessing if the AI system falls under the project archetype of improving, augmenting or automating an existing process}, \textit{translating use cases into goals, metrics and data needs}, \textit{cost-benefit analysis} and \textit{risk analysis}. The objectives outlined in the design phase of the AI life cycle bear similarities to the typical aims of carrying out RE tasks, such as \textit{requirements elicitation}, \textit{requirements analysis}, \textit{requirements specification} and \textit{requirements management}~\cite{berenbach2009software, wiegers2013software}. This led to our current hypothesis that the trustworthy AI development process could benefit from having a systematic RE process and best practices that help in addressing the limitations of existing ethical AI frameworks for the development and operation of trustworthy AI.  

\section{Methodology}

We selected four of the most widely recognized ethical AI guidelines and compared these ethical AI guidelines discussed in \ref{sec:aiguidelines} to each other. We conducted a literature review of grey literature that presents the applicable principles and guidelines that developers and organisations need to adhere to while developing AI systems. This literature review was followed by a qualitative analysis to identify semantic inconsistencies and coverage gaps in the discussed principles for \textbf{RQ1}. We provide a table which resulted from the comparative analysis of the principles focusing on the development of AI systems in the scope of the AIA in Section~\ref{sec:results}.

\begin{figure}[!ht]
    \centering
    \includegraphics[width=.8\linewidth]{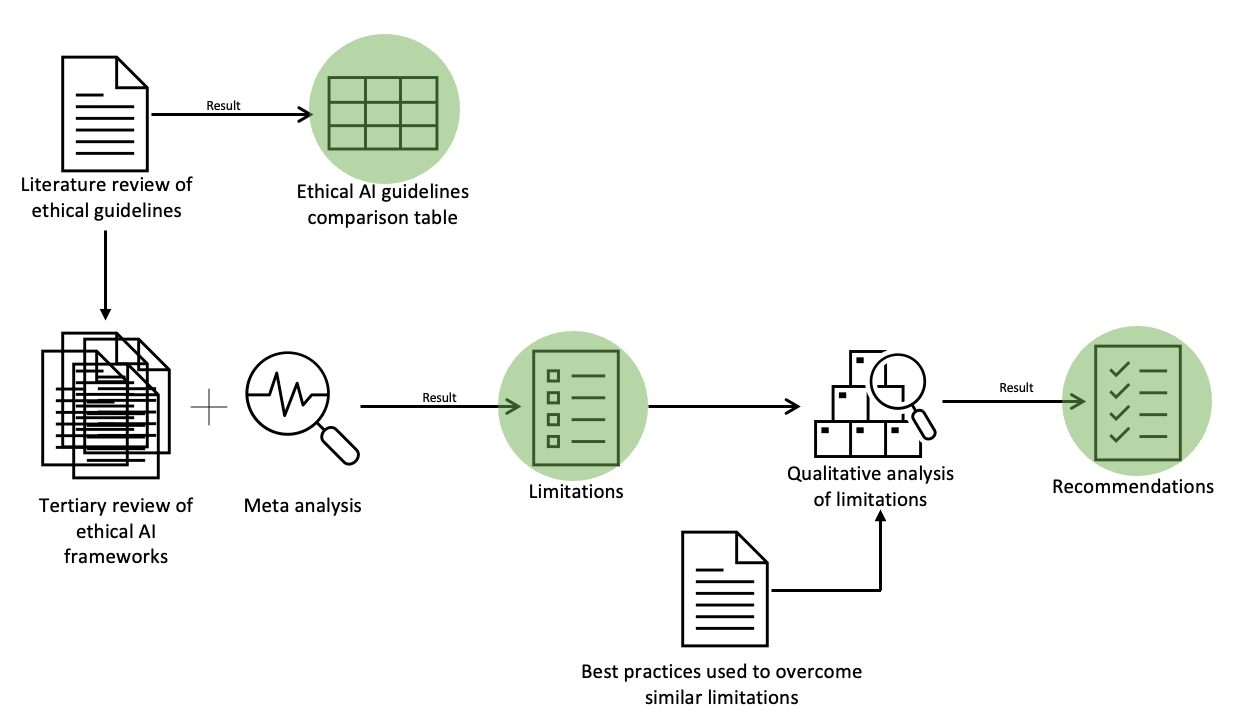}
    \caption{Research methodology employed to answer the three research questions.}
    \label{fig:methodologydiagram}
\end{figure}

We performed a tertiary review (ie., a systematic review of systematic reviews~\cite{article}) of existing frameworks meant to support the development of ethical and responsible AI systems in order to address \textbf{RQ2}. The goal was to understand whether existing ethical AI creation frameworks could help during the development of AI systems that comply with the ethical principles covered in the draft for the AIA which is the most relevant guideline for AI development in the EU~\cite{AIact,aihleg}. In particular, we assessed the applicability of existing ethical AI development frameworks for RE activities. We followed the protocol for a tertiary study of systematic literature reviews and evidence-based guidelines in IT and software engineering~\cite{article} in this study. The results of this study are presented in \ref{sec:frameworkgap}. The process is as follows:

\paragraph{Search process:} We employed an automated search of the following major electronic libraries and databases: Google Scholar, IEEE Xplore, ACM Digital Library, Scopus, ScienceDirect, Web of Science and arXiv by using a set of truncated strings. 

\begin{itemize}
\item Search string formulation: \textit{'Systematic Literature reviews' OR 'Systematic Mapping studies' OR 'Systematic Surveys' OR 'SLR' OR 'SMS' AND 'Frameworks' OR 'Methodology' AND 'Ethical' OR 'Responsible' OR 'Trustworthy' AND 'AI' or 'Artificial Intelligence' OR 'Autonomous Systems'}
\item Inclusion Criteria:
    \begin{itemize}
        \item Type of Studies: Systematic Literature Reviews, Systematic Mapping Studies and Systematic Surveys presenting the analysis of various ethical AI frameworks and methods. 
        \item Type of publications: Conference proceedings and Journals.
        \item Language: English.
        \item Time span: 2019 (since the European Commission (EC) released their initial draft of the AI Act in 2019)-current. 
    \end{itemize}
\item Exclusion Criteria:
    \begin{itemize}
        \item Informal literature surveys.
        \item Papers not subject to peer review.
    \end{itemize}
\end{itemize}

\paragraph{Primary study selection process:} A total of 17,285 results were found using all the possible truncated strings in all the electronic databases mentioned above. IEEE Xplore gave the lowest number of results (13) while Google Scholar gave the most (16,900). Scopus yielded 61 results while ACM digital library gave 72 results, ScienceDirect gave 170 results, arXiv provided 7 papers and Web of Science yielded 62 publications. The number of papers dropped down to a total of 37 after removing the duplicates and filtering the papers based on the title and keyword match. Moreover, Google Scholar was queried for the first 300 results, for subsequent results were no longer relevant, and manually screened by title and abstract. A total of 37 papers matching our inclusion criteria were retrieved and a rigorous filtering process and quality assessment was conducted to select three highly relevant and significant studies. Forward and backward snowballing of those papers resulted in the inclusion of another SLR bringing the total number of included systematic studies to four. These studies provide valuable insights into the development practices of ethical AI, and their inclusion in this study was based on strict criteria aimed at highlighting the most relevant findings.

\paragraph{Quality assessment:} The quality assessment criteria are based on four questions: (i)~Are the review's inclusion and exclusion criteria described and appropriate (i.e, follow a well-accepted systematic study guide)?, (ii)~Is the literature search likely to have covered all relevant studies?, (iii)~Did the reviewers assess the quality/validity of the included studies?, and (iv)~Were the basic data/studies adequately described?

\paragraph{Data collection and analysis:} The data was extracted and analyzed by one researcher and checked by another in the first pass and the roles were exchanged for cross-validation in the second pass to reduce the risk of selection and confirmation bias. A meta-analysis of the extracted data from the identified literature was performed to present the limitations of these frameworks in the development of trustworthy AI. 

\section{Results \& Analysis}\label{sec:results}
\subsection{RQ1: comparative analysis of ethical AI guidelines} \label{sec:guidelinegap}

\begin{table}[]
\begin{tabular}{|l|cccc|c|}
\hline
\multirow{2}{*}{\textbf{Principle}} & \textbf{IEEE-EAD} & \textbf{JPAI} & \textbf{UNESCO} & \textbf{OECD} & \textbf{AIA} \\
& 2017 & 2021 & 2021 & 2022 & 2021 \\
\hline
Accountability & ++ & ++ & ++ & ++ & ++ \\
Accuracy & - & - & - & - & ++ \\
Beneficence & ++ & - & + & ++ & ++ \\
Collaboration & ++ & - & ++ & - & - \\
Competitive Fairness & - & ++ & + & + & + \\
Data Protection & ++ & + & ++ & ++ & ++ \\
Education \& Literacy & ++ & ++ & ++ & - & -  \\
Explainability & + & - & ++ & ++ & ++ \\
Fairness & + & ++ & ++ & ++ & ++ \\
Human Oversight & + & + & ++ & - & ++ \\
Human-centricness & ++ & ++ & + & ++ & ++ \\
Innovation & + & ++ & + & ++ & + \\
No Harm & + & - & ++ & - & ++ \\
Non-discrimination & + & + & ++ & ++ & ++ \\
Privacy & + & ++ & ++ & + & ++ \\
Responsibility & + & - & ++ & ++ & ++ \\
Robustness & + & - & - & ++ & ++ \\
Safety & ++ & + & ++ & ++ & ++ \\
Security & ++ & ++ & ++ & ++ & ++ \\
Sustainability & + & + & ++ & ++ & ++ \\
Transparency & ++ & ++ & ++ & ++ & ++ \\
Well being & ++ & + & ++ & ++ & ++ \\
\hline
\end{tabular}

\caption{Comparison of ethical AI guidelines. The rows are ordered alphabetically while the columns are ordered by the release year in ascending order``++'': the document explicitly defines and provides recommendations for that principle; ``+'': the document either does not clearly define it or provide recommendations for it; ``-'': the document does not discuss this principle at all.} 
\label{tab:guidelinecomparison}
\end{table}

Different guidelines as presented in \autoref{sec:aiguidelines} cover relevant topics for trustworthy AI development but use different definitions for similar concepts. Some of the popular guidelines present or cover ethical principles that are not included in AIA while AIA presents some principles that have not been covered before in other guidelines as presented in \autoref{tab:guidelinecomparison}. 

A major challenge when attempting to compare guidelines is the lack of a unified glossary for ethical-related terms aligned with the terminology of the AIA. Moreover, guidelines describe principles for the development of AI-enabled systems at different abstraction levels. For instance, while AIA intends to promote well-being by imposing requirements on AI systems, the UNESCO guidelines explicitly consider the effect of AI on human health and systems. Meanwhile, other values and principles are not explicitly defined or even have conflicting definitions across guidelines. An example of this is the need for AI to communicate its limitations to the user, which is clearly defined and enforced by the AIA, while other guidelines do not mention it explicitly. For instance, UNESCO guidelines suggest investing in media literacy skills to raise awareness, which could inform final users about the potential limitations of AI. On the other hand, the AIA enforces the need for technical documentation and explicit communication channels for the AI to communicate its limitations with different stakeholders. AIA highlights the need for these principles in AI development: \textit{Privacy}, \textit{No harm}, \textit{Non-discrimination}, \textit{Sustainability}, \textit{Data protection}, \textit{Human oversight}, \textit{Explainability}, \textit{Responsibility}, \textit{Robustness}, \textit{Well-being}, \textit{Beneficence}, and \textit{Accuracy} among others. These principles, crucial for ensuring the ethical development and operation of AI, are not unique to the EU AI Act and are addressed in other guidelines to different extents. Consideration of these principles is becoming increasingly important for policymakers and other stakeholders as the adoption of AI continues to grow in various industries~\cite{10.1145/3430368, kaur2022trustworthy, dwivedi2021artificial} and hence, plays a very important role in innovation and industrial competitiveness.

\subsection{RQ2: limitations of existing AI creation frameworks} \label{sec:frameworkgap}

Different frameworks, as presented in \autoref{sec:frameworks}, have different coverage levels of necessary requirements for AI systems as defined by the AI Act~\cite{AIact}. The state-of-art frameworks' applicability has various limitations. Our meta-analysis of these AI creation frameworks will serve (i) to examine their applicability to the RE process for trustworthy AI and (ii) to understand the limitations and challenges regarding such a process.

While DFV~\cite{crowley2019toward} provides a framework for Requirement Engineers, it may fall short in providing specific guidelines for eliciting ethical requirements that align with the AIA when developing AI systems. Additionally, the responsible development life cycle of AI systems outlined in the framework does not mention requirements explicitly nor facilitates any systematic RE processes. 

Since the four hierarchical levels in the VCIO framework~\cite{vcio}, i.e. values, criteria, indicators, and observables are closely linked to each other, the fulfilment of the higher level depends on the lower level. How the lower-level artefacts of the framework are handled depends largely on how the value is defined. Misinterpretation/ inconsistent definition of a certain value by the organization that does not align with the AIA's own definition can cause problems. The risk matrix does not address how developers and ML engineers can operationalise it to elicit and formulate actionable requirements when developing AI systems. 

 The TAII~\cite{baker2021taii} process for an AI system life cycle lacks specific directions to implement its recommendations in practice, especially during the requirement engineering stage. The level of granularity of this framework in regard to technical implementation is quite low, i.e., provides executive-level organisational guidelines, making it harder for technical stakeholders like developers, data scientists, and product owners to adopt this framework to align the AI system with the EU AI Act guidelines. We find the whole framework and the steps involved focus more on business implementation on how organizations can be ethically compliant on a process level but there is a risk that the compliance will not translate into the product's behaviour. 

Our meta-analysis of the publications included in the tertiary review revealed that certain principles such as explicability and explainability have become more prominent in the development of technical solutions due to their perceived suitability. However, there has not been sufficient emphasis on incorporating mechanisms to evaluate the impact of data processing in ML algorithms on individuals and society. Many of the existing methods are not actionable as they lack practical guidance and require a high level of technical expertise to implement. The typology provided by Morley et al.~\cite{morley2020initial} suggests that no single ethical framework can address all AI-related concerns, and multiple frameworks may need to be applied during the business and use-case development phase for each ethical principle. The lack of actionable resources for implementing ethics in practice is leading to a situation where developers acknowledge ethical principles as a goal but do not pursue them formally~\cite{vakkuri2019implementing}. There is a shortage of processes that specifically focus on eliciting and formulating requirements. In other words, there is a lack of systematic approaches to elicit, analyze, specify, and manage requirements in the development of ethical AI~\cite{Zhou2020}. Proposed technical approaches to address ethical concerns in AI are often narrowly focused and tackle a few ethical issues at a time. This finding refines and confirms previous analyses that have highlighted the limited scope of many technical approaches, which often prioritize only a few ethical issues, such as the explicability and fairness of AI systems~\cite {prem2023ethical}.

Based on our meta-analysis and observations from the background literature review of sample frameworks as presented above, the identified limitations can be summarized as follows:
\begin{itemize}
    \item \textbf{L1}: Limited scope of many technical approaches which are narrowly focused and often prioritize only a few ethical issues at a time~\cite{morley2020initial, prem2023ethical}.
    \item \textbf{L2}: Lack of specific recommendations for AI training data selection past technical aspects, such as quality and integrity of data discussions, or ensuring \textit{explainability} and improving certain \textit{biases}~\cite{prem2023ethical}.
    \item \textbf{L3}: Lack of practical RE-specific best practices and processes targeting the design phase of the ethical AI frameworks for AS development~\cite{morley2020initial, vakkuri2019implementing, Zhou2020}.
    \item \textbf{L4}: Different interpretations of ethical principles exist and it is not straightforwardly clear how to formalise them, leading to inconsistent implementations~\cite{crowley2019toward,vcio}.
    \item \textbf{L5}: Difficult to implement methods and strategies due to vague and high-level solutions~\cite{baker2021taii}. 
    \item \textbf{L6}: Lack of clear indications on how to balance ethical principles, which might not have equal importance in all contexts of AS operation, leading to constrained and biased trade-off mechanisms~\cite{crowley2019toward, morley2020initial}.
\end{itemize}

Finally, to the best of the authors' knowledge, none of the existing ethical AI creation frameworks consider the dynamic nature of applicable ethical AI guidelines, potentially leading to rapidly outdated conformance of the AS to the required regulations \textbf{(L7)}.

\subsection{RQ3: recommendations for trustworthy (er) Autonomous Systems} \label{sec:recommendations}

To tackle the limitations of existing AI creation frameworks ($L_n$) presented above, we propose a number of recommendations ($R_m$). These recommendations aim to foster consistent interpretation of principles to translate them into high-level requirements during the RE phase and enable re-validations of the AI after deployment. Table~\ref{tab:recommendation_mapping_table} maps $L_n$ and $R_m$ to the stages of the Requirements Engineering process that are impacted~\cite{berenbach2009software, wiegers2013software}. It hints at which stage of the requirement engineering process can the recommendations be implemented to address a particular limitation. We expect that the proposed recommendations will also help organizations and researchers from depleting their time and resources to re-iterate and adapt their development frameworks and methodologies every time a newer version of an ethical AI guideline is released in response to the rapidly evolving field of AI to ensure their products' compliance. 

Eight recommendations for different phases of the AI life cycle and adaptations to existing AI creation frameworks are proposed, described, and motivated:

\textbf{(R1): Use a Compositional Architecture Framework:} It is possible to assign a separate cluster of concerns for ethical and trustworthy requirements by using a framework such as the one proposed by Heyn et al.~\cite{heyn2023compositional}, which helps in providing a place for RE activities within the AI development process (L1).

\textbf{(R2): Implement a systematic data selection process:} This fosters ``Trustworthiness by Design'' by bridging the gap between the data selected for the training of the ML model and the requirement engineering process (L2) by establishing traceability mechanisms between system requirements, data requirements, and the completeness criteria~\cite{heyn2023investigation}. 

\textbf{(R3): Develop or adopt RE-specific processes and practices in the design phase:} A major observation from the tertiary review is how RE is simply not mentioned as a process-level necessity. More often than not, it is simply overlooked as a self-evident part of the design phase instead (L3). Considering the importance and value RE provides in fostering trustworthiness in AI from the early stages~\cite{kostova2020interplay,vakkuri2021eccola}, there is a significant advantage in investing resources and effort for the development of RE-specific processes like Ethics-Aware SE~\cite{aydemir2018roadmap} that allows stakeholders to analyse the ethical requirements through a five stage process: 1. \textit{Articulation}, 2. \textit{Specification}, 3. \textit{Implementation}, 4. \textit{Verification} and 5. \textit{Validation} of the requirements.

\textbf{(R4): Utilizing a standard glossary of definitions for the ethical principles across the entire domain of AI development while interpreting and implementing them into the system:} The terminology used to define and represent a certain ethical value differs from guideline to guideline as observed from the results of \textbf{RQ1}. This limitation (L4) can be further aggravated if the already obscure and incongruous definitions of the ethical principles are interpreted differently in an ad-hoc manner. A glossary of ethical principles for AI development through a collaborative effort between a standards organisation like ISO, ANSI or CEN and AI regulatory authorities like the European Commission (EC) or the FTC and NIST in the US will help in the consistent interpretation and translation of principles into high-level requirements by the AI systems development organisations. This will help in ensuring compliance with applicable regulations since the glossary is being developed and approved by the same authorities that oversee the AI systems' compliance.

\textbf{(R5): Develop metrics to evaluate the quality of Trustworthy AI requirements:} Having AI-specific quality evaluation metrics for the requirements will help with verification and validation (L5). This offers a more systematic method to verify non-functional requirements (NFRs) for AI that are hard to evaluate.

\textbf{(R6): Provenance documentation to foster Trustworthiness in AI creation:} Provenance documentation is perceived to be useful in improving fairness, accountability, transparency and explainability~\cite{werder2022establishing}. It can help in keeping track of entities, agents, and activities at each step in the process to support implementing traceability mechanisms to easily track and identify the data origins, and data processing steps~\cite{kale2022provenance}. Provenance documentation can improve transparency and explainability on both the process and product levels.

\textbf{(R7): Establish trade-off patterns for prioritizing the trustworthy AI principles:} Developing and implementing established systematic trade-off patterns for prioritising conflicting ethical principles will help in mitigating the issue of inconsistent and biased prioritisation (L6) of the ethical principles that the AI system needs to embody. The use of absolute scale data to provide unambiguous prioritization data that stakeholders can understand and relate to as proposed by Lindsey et al.~\cite{brodie2011prioritization} is an example that can be used. Another work to consider would be the approach proposed by Ahmed et al.~\cite{saeed2018non}, which showed how the goal model and the system's runtime environment are used for constructing multi-entity bayesian network fragments to reason the trade-off between non-functional requirements for a self-adaptive system. 

\textbf{(R8): Adapt to potential changes in ethical AI guidelines:} Foresight methodologies to establish whether AS should continue abiding by a particular principle in AI guidelines and regular re-applications and re-validations of AI guidelines to AS in operations (L7). Re-validating can help understand when to retire an AS.

\begin{table}[!ht]
\begin{tabular}{|l|l|l|l|l|l|l|l|}
\hline
\textbf{Phase} & \textbf{\textbf{L1}} & \textbf{L2} & \textbf{\textbf{L3}} & \textbf{L4} & \textbf{L5} & \textbf{\textbf{L6}} & \textbf{L7} \\ \hline
\begin{tabular}[c]{@{}l@{}}Requirements\\ elicitation\end{tabular} & R1 & R2 &  &  &  &  &  \\ \hline
\begin{tabular}[c]{@{}l@{}}Requirements\\ analysis\end{tabular} &  &  & R3 & R4 & R5 &  &  \\ \hline
\begin{tabular}[c]{@{}l@{}}Requirements\\ specification\end{tabular} &  &  &  &  & R6 &  &  \\ \hline
\begin{tabular}[c]{@{}l@{}}Requirements \\ management\end{tabular} &  &  &  &  &  & R7 & R8 \\ 
\hline
\end{tabular}
\caption{Map of recommendations ($R_m$) to limitations ($L_n$) and activities of the RE process}
\label{tab:recommendation_mapping_table}
\end{table}

\section{Discussion}

The analysis in Section~\ref{sec:guidelinegap} to address \textbf{RQ1} shows that various ethical AI guidelines from around the world cover different principles from diverse perspectives and abstraction levels. Out of the discussed guidelines, AIA provides the most comprehensive coverage of the addressed and recommended principles~\cite{AIact}. In contrast to most other major guidelines, though, AIA does not emphasise the \textit{Education \& Literacy} principle. This could potentially lead to ASs developed in Europe (i.e., AIA compliant) not being marketable in countries like Japan, which has regulations that consider \textit{Education \& Literacy} as a core principle.

Nevertheless, the lack of standardised definitions, and practices to formalise and interpret these principles, as well as to integrate them into AI systems (L4), makes it difficult to develop ASs that abide by these ethical AI principles. One potential approach to tackle this issue in the development of AI for AS is the establishment of a standard glossary of definitions for the ethical principles as reflected in our recommendation R4. Such practices could prove highly beneficial for practitioners in the field, given the present tendency to focus primarily on issues of safety and security to the exclusion of other considerations (L6).

The demands of the market, as well as changes in societal perspectives, may influence the  regulations governing the development and operation of ASs within the EU (L7). Therefore, it is crucial to provide suitable recommendations for adapting to alterations in the ethical AI guidelines (R8), including the AI Act. This would enable practitioners to design and develop heavy ASs while anticipating potential modifications in the guidelines once such changes come into effect. To ensure adherence to the overarching principles as outlined in the AIA, it is imperative to adopt judicious data selection approaches that can guide practitioners utilizing existing AI development frameworks (R2).

In the development of trustworthy AI for ASs, if similar AI systems have different implementations because of inconsistent interpretations (L4), it is in direct conflict with a major ethical principle like \textit{Fairness}. This highlights the need for appropriate standards (R4), processes (R3), and trade-off mechanisms (R7) to avoid (L3) and (L6) as well. Additionally, useful metrics that evaluate the quality of elicited requirements (R5) are needed to address (L1) and (L5). Current AI creation frameworks lack adequate emphasis on the importance of RE processes and tools (L3), which are necessary for promoting efficient collaboration strategies among development teams. To address this gap, this paper proposes the integration of robust RE processes and tools (R3), which includes a discussion of requirements analysis and documentation based on a set of ethical-AI-specific metrics (R5) for measuring requirement quality. The compositional architecture framework presented by Heyn et al.~\cite{heyn2023compositional} enables the separation of different clusters of concerns, particularly the AI requirements, from other clusters which will aid in portraying RE as a separate process rather than an under-focused sub-process of the design phase. An additional notable consideration in the implementation of AI systems is the quality of data employed in their training and evaluation (L2).

\subsection{Threats to Validity}

We followed Runeson \& Host's guidelines for conducting qualitative research analysis~\cite{runeson2009guidelines} in software engineering to discuss any possible threats to the validity of this research study. One of the potential threats to validity is the selection criteria of the ethical guidelines included in this study to answer \textbf{RQ1}. There is no guarantee that these guidelines are representative of all ethical AI principles either. We employed structured qualitative analysis methods to investigate the extent of the coverage of ethical AI principles in the selected guidelines and a meta-analysis of the included publications of the tertiary review to identify limitations relevant to performing RE. Some of the SLRs included in our tertiary review analysed over 100 ethical AI frameworks. There is still a chance that the presented limitations are not exhaustive. Further investigation to identify more limitations of existing ethical AI development frameworks in performing RE activities is encouraged. 

\section{Conclusion \& Future Work}

The creation and development of trustworthy AI for ASs are a significant concern as their integration into society is rapidly growing, e.g., the penetration rate of ASs is steadily increasing~\cite{fernandez2021trustworthy}. The development of these systems poses multiple challenges that existing ethical AI frameworks cannot address, burdening developers and users with ethical decision-making duties~\cite{prem2023ethical}.

This paper comparatively analyses ethical AI guidelines, in Section~\ref{sec:guidelinegap}, and ethical AI frameworks to understand their limitations for AS development, addressing \textbf{RQ1}. Then, we analysed the limitations of ethical AI development frameworks from an RE perspective for the development of trustworthy AI, answering \textbf{RQ2} as discussed in Section~\ref{sec:frameworkgap}. Our analysis is centred on the AIA, which promotes trustworthiness by design, highlighting the importance of compliance from the early stages. 

Section~\ref{sec:recommendations} moves on to propose appropriate recommendations that address the limitations of current ethical AI frameworks. These recommendations, which constitute the answer to \textbf{RQ3}, benefit the AI development community by providing RE-centric strategies that foster trustworthiness from the early stages of development.

Despite the complexity and rapid evolution of the AI field, we still need to consider establishing a foundation for governing, designing, and implementing trustworthiness into ASs, both on product and process levels. The recommendations also aim at creating a common ground to discuss the ethical implications of AI systems with all the stakeholders. Further research, by carrying out impact analysis and assessment in collaboration with relevant industrial and multi-disciplinary research communities, is needed to further validate the proposed recommendations.

A natural continuation for this work is proposing a RE framework to operationalise and formalise the \textit{trustworthy by design} agenda in the AIA. This framework should be flexible to adapt to regulatory changes and the diverse needs of a society whose reality, values, and needs are constantly changing.

\section{Acknowledgement}

This work was supported by the Vinnova project ASPECT [2021-04347].

\bibliographystyle{ACM-Reference-Format}
\bibliography{main}


\begin{thebibliography}{44}


\ifx \showCODEN    \undefined \def \showCODEN     #1{\unskip}     \fi
\ifx \showDOI      \undefined \def \showDOI       #1{#1}\fi
\ifx \showISBNx    \undefined \def \showISBNx     #1{\unskip}     \fi
\ifx \showISBNxiii \undefined \def \showISBNxiii  #1{\unskip}     \fi
\ifx \showISSN     \undefined \def \showISSN      #1{\unskip}     \fi
\ifx \showLCCN     \undefined \def \showLCCN      #1{\unskip}     \fi
\ifx \shownote     \undefined \def \shownote      #1{#1}          \fi
\ifx \showarticletitle \undefined \def \showarticletitle #1{#1}   \fi
\ifx \showURL      \undefined \def \showURL       {\relax}        \fi
\providecommand\bibfield[2]{#2}
\providecommand\bibinfo[2]{#2}
\providecommand\natexlab[1]{#1}
\providecommand\showeprint[2][]{arXiv:#2}

\bibitem[chi([n.\,d.])]%
        {chinaAI}
 \bibinfo{year}{[n.\,d.]}\natexlab{}.
\newblock \bibinfo{title}{Translation: Measures for the Management of
  Generative Artificial Intelligence Services (Draft for Comment) – April
  2023}.
\newblock
\newblock
\urldef\tempurl%
\url{https://digichina.stanford.edu/work/translation-measures-for-the-management-of-generative-artificial-intelligence-services-draft-for-comment-april-2023/}
\showURL{%
\tempurl}


\bibitem[Amaral et~al\mbox{.}(2021)]%
        {amaral2021trustworthiness}
\bibfield{author}{\bibinfo{person}{Glenda Amaral}, \bibinfo{person}{Renata
  Guizzardi}, \bibinfo{person}{Giancarlo Guizzardi}, {and}
  \bibinfo{person}{John Mylopoulos}.} \bibinfo{year}{2021}\natexlab{}.
\newblock \showarticletitle{Trustworthiness requirements: the pix case study}.
  In \bibinfo{booktitle}{\emph{Conceptual Modeling: 40th International
  Conference, ER 2021, Virtual Event, October 18--21, 2021, Proceedings 40}}.
  Springer, \bibinfo{pages}{257--267}.
\newblock


\bibitem[Aydemir and Dalpiaz(2018)]%
        {aydemir2018roadmap}
\bibfield{author}{\bibinfo{person}{Fatma~Ba{\c{s}}ak Aydemir} {and}
  \bibinfo{person}{Fabiano Dalpiaz}.} \bibinfo{year}{2018}\natexlab{}.
\newblock \showarticletitle{A roadmap for ethics-aware software engineering}.
  In \bibinfo{booktitle}{\emph{Proceedings of the International Workshop on
  Software Fairness}}. \bibinfo{pages}{15--21}.
\newblock


\bibitem[Baker-Brunnbauer(2021)]%
        {baker2021taii}
\bibfield{author}{\bibinfo{person}{Josef Baker-Brunnbauer}.}
  \bibinfo{year}{2021}\natexlab{}.
\newblock \showarticletitle{TAII framework for trustworthy AI systems}.
\newblock \bibinfo{journal}{\emph{Baker-Brunnbauer, J.(2021). TAII Framework
  for Trustworthy AI Systems. ROBONOMICS: The Journal of the Automated
  Economy}}  \bibinfo{volume}{2} (\bibinfo{year}{2021}), \bibinfo{pages}{17}.
\newblock


\bibitem[Berenbach et~al\mbox{.}(2009)]%
        {berenbach2009software}
\bibfield{author}{\bibinfo{person}{Brian Berenbach}, \bibinfo{person}{Daniel~J
  Paulish}, \bibinfo{person}{Juergen Kazmeier}, {and} \bibinfo{person}{Arnold
  Rudorfer}.} \bibinfo{year}{2009}\natexlab{}.
\newblock \bibinfo{booktitle}{\emph{Software \& systems requirements
  engineering: in practice}}.
\newblock \bibinfo{publisher}{McGraw-Hill Education}.
\newblock


\bibitem[Brodie and Woodman(2011)]%
        {brodie2011prioritization}
\bibfield{author}{\bibinfo{person}{Lindsey Brodie} {and} \bibinfo{person}{Mark
  Woodman}.} \bibinfo{year}{2011}\natexlab{}.
\newblock \showarticletitle{Prioritization of stakeholder value using metrics}.
  In \bibinfo{booktitle}{\emph{Evaluation of Novel Approaches to Software
  Engineering: 5th International Conference, ENASE 2010, Athens, Greece, July
  22-24, 2010, Revised Selected Papers 5}}. Springer, \bibinfo{pages}{74--88}.
\newblock


\bibitem[Campbell and Glass(2000)]%
        {campbell2000legal}
\bibfield{author}{\bibinfo{person}{Angela Campbell} {and}
  \bibinfo{person}{Kathleen~Cranley Glass}.} \bibinfo{year}{2000}\natexlab{}.
\newblock \showarticletitle{The legal status of clinical and ethics policies,
  codes, and guidelines in medical practice and research}.
\newblock \bibinfo{journal}{\emph{McGill LJ}}  \bibinfo{volume}{46}
  (\bibinfo{year}{2000}), \bibinfo{pages}{473}.
\newblock


\bibitem[Canca(2020)]%
        {10.1145/3430368}
\bibfield{author}{\bibinfo{person}{Cansu Canca}.}
  \bibinfo{year}{2020}\natexlab{}.
\newblock \showarticletitle{Operationalizing AI Ethics Principles}.
\newblock \bibinfo{journal}{\emph{Commun. ACM}} \bibinfo{volume}{63},
  \bibinfo{number}{12} (\bibinfo{date}{nov} \bibinfo{year}{2020}),
  \bibinfo{pages}{18–21}.
\newblock
\showISSN{0001-0782}
\urldef\tempurl%
\url{https://doi.org/10.1145/3430368}
\showDOI{\tempurl}


\bibitem[Chesterman(2020)]%
        {chesterman2020artificial}
\bibfield{author}{\bibinfo{person}{Simon Chesterman}.}
  \bibinfo{year}{2020}\natexlab{}.
\newblock \showarticletitle{Artificial intelligence and the problem of
  autonomy}.
\newblock \bibinfo{journal}{\emph{Notre Dame J. on Emerging Tech.}}
  \bibinfo{volume}{1} (\bibinfo{year}{2020}), \bibinfo{pages}{210}.
\newblock


\bibitem[Crowley et~al\mbox{.}(2019)]%
        {crowley2019toward}
\bibfield{author}{\bibinfo{person}{James Crowley}, \bibinfo{person}{AP
  OrSullivan}, \bibinfo{person}{Andrzej Nowak}, \bibinfo{person}{Catholijn
  Jonker}, \bibinfo{person}{Dino Pedreschi}, \bibinfo{person}{Fosca Giannotti},
  {and} \bibinfo{person}{Y Rogers}.} \bibinfo{year}{2019}\natexlab{}.
\newblock \showarticletitle{Toward ai systems that augment and empower humans
  by understanding us, our society and the world around us}.
\newblock \bibinfo{journal}{\emph{Report of}}  \bibinfo{volume}{761758}
  (\bibinfo{year}{2019}), \bibinfo{pages}{1--32}.
\newblock


\bibitem[Dechesne(2020)]%
        {9218169}
\bibfield{author}{\bibinfo{person}{Francien Dechesne}.}
  \bibinfo{year}{2020}\natexlab{}.
\newblock \showarticletitle{Requirements engineering for moral considerations
  in algorithmic systems : RE’20 Conference Keynote}. In
  \bibinfo{booktitle}{\emph{2020 IEEE 28th International Requirements
  Engineering Conference (RE)}}. \bibinfo{pages}{1--2}.
\newblock
\urldef\tempurl%
\url{https://doi.org/10.1109/RE48521.2020.00010}
\showDOI{\tempurl}


\bibitem[Dignum and Confidential(2019)]%
        {dignum2019humane}
\bibfield{author}{\bibinfo{person}{Virginia Dignum} {and} \bibinfo{person}{CO
  Confidential}.} \bibinfo{year}{2019}\natexlab{}.
\newblock \showarticletitle{HumanE AI}.
\newblock  (\bibinfo{year}{2019}).
\newblock


\bibitem[Dwivedi et~al\mbox{.}(2021)]%
        {dwivedi2021artificial}
\bibfield{author}{\bibinfo{person}{Yogesh~K Dwivedi}, \bibinfo{person}{Laurie
  Hughes}, \bibinfo{person}{Elvira Ismagilova}, \bibinfo{person}{Gert Aarts},
  \bibinfo{person}{Crispin Coombs}, \bibinfo{person}{Tom Crick},
  \bibinfo{person}{Yanqing Duan}, \bibinfo{person}{Rohita Dwivedi},
  \bibinfo{person}{John Edwards}, \bibinfo{person}{Aled Eirug},
  {et~al\mbox{.}}} \bibinfo{year}{2021}\natexlab{}.
\newblock \showarticletitle{Artificial Intelligence (AI): Multidisciplinary
  perspectives on emerging challenges, opportunities, and agenda for research,
  practice and policy}.
\newblock \bibinfo{journal}{\emph{International Journal of Information
  Management}}  \bibinfo{volume}{57} (\bibinfo{year}{2021}),
  \bibinfo{pages}{101994}.
\newblock


\bibitem[et~al(2020)]%
        {vcio}
\bibfield{author}{\bibinfo{person}{Hallensleben et al}.}
  \bibinfo{year}{2020}\natexlab{}.
\newblock \bibinfo{booktitle}{\emph{Artificial Intelligence Ethics Impact Group
  AIEI Group. From Principles to Practice. An Interdisciplinary Framework to
  Operationalise AI Ethics}}.
\newblock
\urldef\tempurl%
\url{https://www.ai-ethics-impact.org/resource/blob/1961130/c6db9894ee73aefa489d6249f5ee2b9f/aieig---report---download-hb-data.pdf}
\showURL{%
\tempurl}


\bibitem[European~Commission and Technology(2021)]%
        {AIact}
\bibfield{author}{\bibinfo{person}{Content European~Commission,
  Directorate-General for Communications~Networks} {and}
  \bibinfo{person}{Technology}.} \bibinfo{year}{2021}\natexlab{}.
\newblock \bibinfo{booktitle}{\emph{EUR-Lex - 52021PC0206 - EN - EUR-Lex}}.
\newblock
\urldef\tempurl%
\url{https://eur-lex.europa.eu/legal-content/EN/TXT/?uri=CELEX:52021PC0206}
\showURL{%
\tempurl}


\bibitem[Fern{\'a}ndez~Llorca and G{\'o}mez(2021)]%
        {fernandez2021trustworthy}
\bibfield{author}{\bibinfo{person}{David Fern{\'a}ndez~Llorca} {and}
  \bibinfo{person}{Emilia G{\'o}mez}.} \bibinfo{year}{2021}\natexlab{}.
\newblock \showarticletitle{Trustworthy autonomous vehicles}.
\newblock \bibinfo{journal}{\emph{Publications Office of the European Union,
  Luxembourg,, EUR}}  \bibinfo{volume}{30942} (\bibinfo{year}{2021}).
\newblock
\urldef\tempurl%
\url{https://doi.org/doi/10.2760/120385}
\showDOI{\tempurl}


\bibitem[Floridi et~al\mbox{.}(2022)]%
        {floridi2022capai}
\bibfield{author}{\bibinfo{person}{Luciano Floridi}, \bibinfo{person}{Matthias
  Holweg}, \bibinfo{person}{Mariarosaria Taddeo}, \bibinfo{person}{Javier
  Amaya~Silva}, \bibinfo{person}{Jakob M{\"o}kander}, {and}
  \bibinfo{person}{Yuni Wen}.} \bibinfo{year}{2022}\natexlab{}.
\newblock \showarticletitle{CapAI-A Procedure for Conducting Conformity
  Assessment of AI Systems in Line with the EU Artificial Intelligence Act}.
\newblock \bibinfo{journal}{\emph{Available at SSRN 4064091}}
  (\bibinfo{year}{2022}).
\newblock


\bibitem[Heyn et~al\mbox{.}(2023b)]%
        {heyn2023investigation}
\bibfield{author}{\bibinfo{person}{Hans-Martin Heyn}, \bibinfo{person}{Eric
  Knauss}, \bibinfo{person}{Iswarya Malleswaran}, {and}
  \bibinfo{person}{Shruthi Dinakaran}.} \bibinfo{year}{2023}\natexlab{b}.
\newblock \showarticletitle{An investigation of challenges encountered when
  specifying training data and runtime monitors for safety critical ML
  applications}.
\newblock \bibinfo{journal}{\emph{arXiv preprint arXiv:2301.13476}}
  (\bibinfo{year}{2023}).
\newblock


\bibitem[Heyn et~al\mbox{.}(2023a)]%
        {heyn2023compositional}
\bibfield{author}{\bibinfo{person}{Hans-Martin Heyn}, \bibinfo{person}{Eric
  Knauss}, {and} \bibinfo{person}{Patrizio Pelliccione}.}
  \bibinfo{year}{2023}\natexlab{a}.
\newblock \showarticletitle{A compositional approach to creating architecture
  frameworks with an application to distributed AI systems}.
\newblock \bibinfo{journal}{\emph{Journal of Systems and Software}}
  (\bibinfo{year}{2023}), \bibinfo{pages}{111604}.
\newblock


\bibitem[IEEE(2021)]%
        {ieee}
\bibfield{author}{\bibinfo{person}{IEEE}.} \bibinfo{year}{2021}\natexlab{}.
\newblock \bibinfo{booktitle}{\emph{The IEEE Global Initiative on Ethics of
  Autonomous and Intelligent Systems. Ethically Aligned Design: A Vision for
  Prioritizing Human Well-being with Autonomous and Intelligent Systems,
  Version 2. IEEE, 2017}}.
\newblock \bibinfo{publisher}{IEEE}.
\newblock
\urldef\tempurl%
\url{http://standards.ieee.org/develop/indconn/ec/autonomous_systems.html}
\showURL{%
\tempurl}


\bibitem[Implemented(2021)]%
        {JapanAI}
\bibfield{author}{\bibinfo{person}{Expert Group On How AI Principles Should~Be
  Implemented}.} \bibinfo{year}{2021}\natexlab{}.
\newblock \showarticletitle{AI Governance in Japan Ver. 1.1}.
\newblock  (\bibinfo{year}{2021}).
\newblock
\urldef\tempurl%
\url{https://www.meti.go.jp/english/press/2020/0616_001.html.}
\showURL{%
\tempurl}


\bibitem[Iyer(2021)]%
        {iyer2021ai}
\bibfield{author}{\bibinfo{person}{Lakshmi~Shankar Iyer}.}
  \bibinfo{year}{2021}\natexlab{}.
\newblock \showarticletitle{AI enabled applications towards intelligent
  transportation}.
\newblock \bibinfo{journal}{\emph{Transportation Engineering}}
  \bibinfo{volume}{5} (\bibinfo{year}{2021}), \bibinfo{pages}{100083}.
\newblock


\bibitem[Jobin et~al\mbox{.}(2019)]%
        {jobin2019global}
\bibfield{author}{\bibinfo{person}{Anna Jobin}, \bibinfo{person}{Marcello
  Ienca}, {and} \bibinfo{person}{Effy Vayena}.}
  \bibinfo{year}{2019}\natexlab{}.
\newblock \showarticletitle{The global landscape of AI ethics guidelines}.
\newblock \bibinfo{journal}{\emph{Nature Machine Intelligence}}
  \bibinfo{volume}{1}, \bibinfo{number}{9} (\bibinfo{year}{2019}),
  \bibinfo{pages}{389--399}.
\newblock


\bibitem[Kale et~al\mbox{.}(2022)]%
        {kale2022provenance}
\bibfield{author}{\bibinfo{person}{Amruta Kale}, \bibinfo{person}{Tin Nguyen},
  \bibinfo{person}{Frederick~C Harris~Jr}, \bibinfo{person}{Chenhao Li},
  \bibinfo{person}{Jiyin Zhang}, {and} \bibinfo{person}{Xiaogang Ma}.}
  \bibinfo{year}{2022}\natexlab{}.
\newblock \showarticletitle{Provenance documentation to enable explainable and
  trustworthy AI: A literature review}.
\newblock \bibinfo{journal}{\emph{Data Intelligence}} (\bibinfo{year}{2022}),
  \bibinfo{pages}{1--41}.
\newblock


\bibitem[Kaur et~al\mbox{.}(2022)]%
        {kaur2022trustworthy}
\bibfield{author}{\bibinfo{person}{Davinder Kaur}, \bibinfo{person}{Suleyman
  Uslu}, \bibinfo{person}{Kaley~J Rittichier}, {and} \bibinfo{person}{Arjan
  Durresi}.} \bibinfo{year}{2022}\natexlab{}.
\newblock \showarticletitle{Trustworthy artificial intelligence: a review}.
\newblock \bibinfo{journal}{\emph{ACM Computing Surveys (CSUR)}}
  \bibinfo{volume}{55}, \bibinfo{number}{2} (\bibinfo{year}{2022}),
  \bibinfo{pages}{1--38}.
\newblock


\bibitem[Kitchenham and Charters(2007)]%
        {article}
\bibfield{author}{\bibinfo{person}{Barbara Kitchenham} {and}
  \bibinfo{person}{Stuart Charters}.} \bibinfo{year}{2007}\natexlab{}.
\newblock \showarticletitle{Guidelines for performing Systematic Literature
  Reviews in Software Engineering}.
\newblock   \bibinfo{volume}{2} (\bibinfo{date}{01} \bibinfo{year}{2007}).
\newblock


\bibitem[Kostova et~al\mbox{.}(2020)]%
        {kostova2020interplay}
\bibfield{author}{\bibinfo{person}{Blagovesta Kostova}, \bibinfo{person}{Seda
  Gurses}, {and} \bibinfo{person}{Alain Wegmann}.}
  \bibinfo{year}{2020}\natexlab{}.
\newblock \showarticletitle{On the Interplay between Requirements, Engineering,
  and Artificial Intelligence.}. In \bibinfo{booktitle}{\emph{REFSQ
  Workshops}}.
\newblock


\bibitem[Mittelstadt(2019)]%
        {mittelstadt2019principles}
\bibfield{author}{\bibinfo{person}{Brent Mittelstadt}.}
  \bibinfo{year}{2019}\natexlab{}.
\newblock \showarticletitle{Principles alone cannot guarantee ethical AI}.
\newblock \bibinfo{journal}{\emph{Nature Machine Intelligence}}
  \bibinfo{volume}{1}, \bibinfo{number}{11} (\bibinfo{year}{2019}),
  \bibinfo{pages}{501--507}.
\newblock


\bibitem[Morley et~al\mbox{.}(2020)]%
        {morley2020initial}
\bibfield{author}{\bibinfo{person}{Jessica Morley}, \bibinfo{person}{Luciano
  Floridi}, \bibinfo{person}{Libby Kinsey}, {and} \bibinfo{person}{Anat
  Elhalal}.} \bibinfo{year}{2020}\natexlab{}.
\newblock \showarticletitle{From what to how: an initial review of publicly
  available AI ethics tools, methods and research to translate principles into
  practices}.
\newblock \bibinfo{journal}{\emph{Science and engineering ethics}}
  \bibinfo{volume}{26}, \bibinfo{number}{4} (\bibinfo{year}{2020}),
  \bibinfo{pages}{2141--2168}.
\newblock


\bibitem[OECD(2019)]%
        {OECD}
\bibfield{author}{\bibinfo{person}{OECD}.} \bibinfo{year}{2019}\natexlab{}.
\newblock \bibinfo{booktitle}{\emph{Artificial Intelligence in Society}}.
\newblock \bibinfo{publisher}{OECD}.
\newblock
\showISBNx{9789264582545}
\urldef\tempurl%
\url{https://doi.org/10.1787/eedfee77-en}
\showDOI{\tempurl}


\bibitem[on~Artificial~Intelligence(2019)]%
        {aihleg}
\bibfield{author}{\bibinfo{person}{High-Level Expert~Group on
  Artificial~Intelligence}.} \bibinfo{year}{2019}\natexlab{}.
\newblock \bibinfo{booktitle}{\emph{Ethics Guidelines for Trustworthy
  Artificial Intelligence (AI)}}.
\newblock
\urldef\tempurl%
\url{www.aepd.es/sites/default/files/2019-12/ai-ethics-guidelines.pdf}
\showURL{%
\tempurl}


\bibitem[Prem(2023)]%
        {prem2023ethical}
\bibfield{author}{\bibinfo{person}{Erich Prem}.}
  \bibinfo{year}{2023}\natexlab{}.
\newblock \showarticletitle{From ethical AI frameworks to tools: a review of
  approaches}.
\newblock \bibinfo{journal}{\emph{AI and Ethics}} (\bibinfo{year}{2023}),
  \bibinfo{pages}{1--18}.
\newblock


\bibitem[research and innovation(2019)]%
        {RRI}
\bibfield{author}{\bibinfo{person}{UK research} {and}
  \bibinfo{person}{innovation}.} \bibinfo{year}{2019}\natexlab{}.
\newblock \bibinfo{booktitle}{\emph{responsible research and innovation (RRI)
  framework}}.
\newblock
\urldef\tempurl%
\url{https://www.ukri.org/councils/epsrc/guidance-for-applicants/what-to-include-in-your-proposal/health-technologies-impact-and-translation-toolkit/research-integrity-in-healthcare-technologies/responsible-research-and-innovation/}
\showURL{%
\tempurl}


\bibitem[Runeson and H{\"o}st(2009)]%
        {runeson2009guidelines}
\bibfield{author}{\bibinfo{person}{Per Runeson} {and} \bibinfo{person}{Martin
  H{\"o}st}.} \bibinfo{year}{2009}\natexlab{}.
\newblock \showarticletitle{Guidelines for conducting and reporting case study
  research in software engineering}.
\newblock \bibinfo{journal}{\emph{Empirical software engineering}}
  \bibinfo{volume}{14} (\bibinfo{year}{2009}), \bibinfo{pages}{131--164}.
\newblock


\bibitem[Saeed and Lee(2018)]%
        {saeed2018non}
\bibfield{author}{\bibinfo{person}{Ahmed Abdo~Ali Saeed} {and}
  \bibinfo{person}{Seok-Won Lee}.} \bibinfo{year}{2018}\natexlab{}.
\newblock \showarticletitle{Non-functional requirements trade-off in
  self-adaptive systems}. In \bibinfo{booktitle}{\emph{2018 4th International
  Workshop on Requirements Engineering for Self-Adaptive, Collaborative, and
  Cyber Physical Systems (RESACS)}}. IEEE, \bibinfo{pages}{9--15}.
\newblock


\bibitem[Sifakis and Harel(2022)]%
        {sifakis2022trustworthy}
\bibfield{author}{\bibinfo{person}{Joseph Sifakis} {and} \bibinfo{person}{David
  Harel}.} \bibinfo{year}{2022}\natexlab{}.
\newblock \showarticletitle{Trustworthy Autonomous System Development}.
\newblock \bibinfo{journal}{\emph{ACM Transactions on Embedded Computing
  Systems}} (\bibinfo{year}{2022}).
\newblock


\bibitem[Sossin and Smith(2002)]%
        {sossin2002hard}
\bibfield{author}{\bibinfo{person}{Lorne Sossin} {and}
  \bibinfo{person}{Charles~W Smith}.} \bibinfo{year}{2002}\natexlab{}.
\newblock \showarticletitle{Hard choices and soft law: ethical codes, policy
  guidelines and the role of the courts in regulating government}.
\newblock \bibinfo{journal}{\emph{Alta. L. Rev.}}  \bibinfo{volume}{40}
  (\bibinfo{year}{2002}), \bibinfo{pages}{867}.
\newblock


\bibitem[Thiebes et~al\mbox{.}(2021)]%
        {thiebes2021trustworthy}
\bibfield{author}{\bibinfo{person}{Scott Thiebes}, \bibinfo{person}{Sebastian
  Lins}, {and} \bibinfo{person}{Ali Sunyaev}.} \bibinfo{year}{2021}\natexlab{}.
\newblock \showarticletitle{Trustworthy artificial intelligence}.
\newblock \bibinfo{journal}{\emph{Electronic Markets}}  \bibinfo{volume}{31}
  (\bibinfo{year}{2021}), \bibinfo{pages}{447--464}.
\newblock


\bibitem[UNESCO(2022)]%
        {unesco}
\bibfield{author}{\bibinfo{person}{UNESCO}.} \bibinfo{year}{2022}\natexlab{}.
\newblock \bibinfo{booktitle}{\emph{Recommendation on the Ethics of Artificial
  Intelligence}}.
\newblock
\urldef\tempurl%
\url{https://unesdoc.unesco.org/ark:/48223/pf0000381137}
\showURL{%
\tempurl}


\bibitem[Vakkuri et~al\mbox{.}(2019)]%
        {vakkuri2019implementing}
\bibfield{author}{\bibinfo{person}{Ville Vakkuri},
  \bibinfo{person}{Kai-Kristian Kemell}, {and} \bibinfo{person}{Pekka
  Abrahamsson}.} \bibinfo{year}{2019}\natexlab{}.
\newblock \showarticletitle{Implementing ethics in AI: initial results of an
  industrial multiple case study}. In \bibinfo{booktitle}{\emph{Product-Focused
  Software Process Improvement: 20th International Conference, PROFES 2019,
  Barcelona, Spain, November 27--29, 2019, Proceedings 20}}. Springer,
  \bibinfo{pages}{331--338}.
\newblock


\bibitem[Vakkuri et~al\mbox{.}(2021)]%
        {vakkuri2021eccola}
\bibfield{author}{\bibinfo{person}{Ville Vakkuri},
  \bibinfo{person}{Kai-Kristian Kemell}, \bibinfo{person}{Marianna Jantunen},
  \bibinfo{person}{Erika Halme}, {and} \bibinfo{person}{Pekka Abrahamsson}.}
  \bibinfo{year}{2021}\natexlab{}.
\newblock \showarticletitle{ECCOLA—A method for implementing ethically
  aligned AI systems}.
\newblock \bibinfo{journal}{\emph{Journal of Systems and Software}}
  \bibinfo{volume}{182} (\bibinfo{year}{2021}), \bibinfo{pages}{111067}.
\newblock


\bibitem[Werder et~al\mbox{.}(2022)]%
        {werder2022establishing}
\bibfield{author}{\bibinfo{person}{Karl Werder},
  \bibinfo{person}{Balasubramaniam Ramesh}, {and} \bibinfo{person}{Rongen
  Zhang}.} \bibinfo{year}{2022}\natexlab{}.
\newblock \showarticletitle{Establishing data provenance for responsible
  artificial intelligence systems}.
\newblock \bibinfo{journal}{\emph{ACM Transactions on Management Information
  Systems (TMIS)}} \bibinfo{volume}{13}, \bibinfo{number}{2}
  (\bibinfo{year}{2022}), \bibinfo{pages}{1--23}.
\newblock


\bibitem[Wiegers and Beatty(2013)]%
        {wiegers2013software}
\bibfield{author}{\bibinfo{person}{Karl Wiegers} {and} \bibinfo{person}{Joy
  Beatty}.} \bibinfo{year}{2013}\natexlab{}.
\newblock \bibinfo{booktitle}{\emph{Software requirements}}.
\newblock \bibinfo{publisher}{Pearson Education}.
\newblock


\bibitem[Zhou et~al\mbox{.}(2020)]%
        {Zhou2020}
\bibfield{author}{\bibinfo{person}{Jianlong Zhou}, \bibinfo{person}{Fang Chen},
  \bibinfo{person}{Adam Berry}, \bibinfo{person}{Mike Reed},
  \bibinfo{person}{Shujia Zhang}, {and} \bibinfo{person}{Siobhan Savage}.}
  \bibinfo{year}{2020}\natexlab{}.
\newblock \showarticletitle{A Survey on Ethical Principles of AI and
  Implementations}.
\newblock \bibinfo{journal}{\emph{2020 IEEE Symposium Series on Computational
  Intelligence, SSCI 2020}} (\bibinfo{date}{12} \bibinfo{year}{2020}),
  \bibinfo{pages}{3010--3017}.
\newblock
\showISBNx{9781728125473}
\urldef\tempurl%
\url{https://doi.org/10.1109/SSCI47803.2020.9308437}
\showDOI{\tempurl}


\end{thebibliography}

\end{document}